\documentclass[epj]{svjour}
\usepackage{graphics,amsfonts,amsmath,epsfig}
\newcommand*{\bw}{\begin{widetext}}
\newcommand*{\ew}{\end{widetext}}
\newcommand*{\be}{\begin{equation}}
\newcommand*{\ee}{\end{equation}}
\newcommand*{\cc}[1]{\bar{#1}}
\newcommand*{\re}[1]{\operatorname{Re}{#1}}
\newcommand*{\im}[1]{\operatorname{Im}{#1}}

\newcommand*{\bto}{\mbox{$\stackrel{\mathcal{B}}{\longrightarrow}$}}
\makeatletter

\newcommand{\Rmnum}[1]{\expandafter\@slowromancap\romannumeral #1@}
\makeatother
\newcommand{\RI}{\mathrm{\Rmnum{1}}}
\newcommand{\RII}{\mathrm{\Rmnum{2}}}
\newcommand{\RIII}{\mathrm{\Rmnum{3}}}
\begin{document}
%

\title{Generic example of algebraic bosonisation}
\author{Katarina~Ro\v zman \and D.~K.~Sunko
\thanks{\emph{Email:} dks@phy.hr}%
}                     
\institute{Department of Physics, Faculty of Science, University of Zagreb, Bijeni\v cka cesta 32, HR-10000 Zagreb, Croatia.}
\date{Received: date / Revised version: date}
%
\abstract{
Two identical non-interacting fermions in a three-dimensional harmonic oscillator well are bosonised exactly according to a recently developed general algebraic scheme. Rotational invariance is taken into account within the scheme for the first time. The example is generic for the excitation spectra of finite systems, in particular for the appearance of bands in spectra. A connection to the formalism of the fractional quantum Hall effect is pointed out.
}
\maketitle

\section{Introduction}

Bosonisation of Fermi systems was initiated by F.~Bloch~\cite{Bloch34}, who postulated that their excitations could be represented by density waves. These were later codified in the versatile concept of particle-hole excitations~\cite{Pines66}. Yet the original idea could literally be carried out only in one dimension~\cite{Tomonaga50}.

The bosonisation problem has two aspects, kinematic and dynamic. The dynamic problem, which we do not treat here, is whether low-lying states of a concrete Fermi system may be approximately described by effective bosonic excitations. The kinematic problem is how to rewrite the Hilbert space of fermions exactly in terms of bosonic excitations, if possible. It was historically much investigated in the context of nuclear physics by various coupling schemes, in which an approximate bosonic excitation is constructed from an even number of fermions~\cite{Janssen74,Blaizot78-2}. More recently, few-electron quantum dots have attracted considerable interest as possible electronic or computational devices~\cite{Warburton98,Kyriakidis07}. The underlying physics is formally similar to that of protons in a nucleus: fermions with a repulsive Coulomb interaction are confined by a well potential, which is a harmonic oscillator to zeroth order~\cite{Kalliakos09}. Here we exploit the general structure of many-body Hilbert space to rewrite the oscillator basis in a qualitatively new way. The salient features of the approach can be observed already on the case of two particles, which we use as a generic example for finite systems.

It has recently become apparent~\cite{Sunko16-1,Sunko16-2} that many-body Hilbert space is a finitely generated free module, such that any antisymmetric wave function of $N$ particles may be written
\be
\Psi=\sum_{i=1}^D\Phi_i\Psi_i,
\label{scheme}
\ee
where $D=N!^{d-1}$ is the dimension of the module, with $N$ the number of particles and $d$ the space dimension. This would be a vector space if the coefficients $\Phi_i$ were $c$-numbers. Instead, they are symmetric functions describing bosonic excitations of the antisymmetric functions $\Psi_i$, called \emph{shapes}, which generate the whole Hilbert space by Eq.~\eqref{scheme}, acting as vacua to the excitations $\Phi_i$. A particular basis for these bosonic excitations is called \emph{Euler bosons}. Notably, Euler bosons have no vacuum energy, which is contained in the fermionic shapes. Furthermore, Eq.~\eqref{scheme} shows why Bloch's program could not succeed for $d>1$ in general. Bosonisation has a multiplicative excitation structure, which corresponds to Eq.~\eqref{scheme} only for $d=1$, when the only shape is the well-known ground-state Slater determinant. For $d>1$, the structure is \emph{additive} over different vacua, where one vacuum cannot be obtained as an Euler-boson excitation of another --- they behave like prime numbers --- because shapes are precisely the orthogonal complement to states which contain Euler bosons~\cite{Sunko16-1}. Finally, the algebraic structure gives a generic reason for bands in spectra, in which the various shapes $\Psi_i$ appear as band-heads, excited by the Euler bosons within each band. In this way, the kinematic bosonisation problem is solved exactly as far as possible, while the ``impossible'' part is explicitly given by the shapes, which can be obtained algorithmically in closed form.

In the present work, we give a textbook elaboration of the scheme~\eqref{scheme} for the simplest case: two identical non-interacting fermions in a harmonic potential, corresponding, for example, to two polarized electrons in a quantum dot. It provides a template to understand the shape structure in a well-known setting. The space of the first two excited shells is rewritten in Euler bosons and shapes in an explicitly rotationally invariant form, which extends previous work in the Cartesian basis~\cite{Sunko16-1}.

\section{Technique}

The Bargmann transform~\cite{Bargmann61} reads
\be
\mathcal{B}[f](t)=\frac{1}{\pi^{1/4}}
\int_{\mathbb{R}}dx\,
e^{-\frac{1}{2}\left(t^2+x^2\right)+xt\sqrt{2}}f(x)
\equiv F(t).
\ee
Here $f\in\mathrm{L}^2(\mathbb{R})$ and $F\in\mathrm{F}(\mathbb{C})$, the \emph{Bargmann space} of entire functions $F:\mathbb{C}\to\mathbb{C}$ such that
\be
\int_{\mathbb{C}}|F(t)|^2d\lambda(t)<\infty,
\ee
where
\be
d\lambda(t)=\frac{1}{\pi}e^{-|t|^2}d\re{t}\,d\im{t},\quad
\int_{\mathbb{C}}d\lambda(t)=1.
\ee
The inverse Bargmann transform is then
\be
\mathcal{B}^{-1}[F](x)=\frac{1}{\pi^{1/4}}
\int_{\mathbb{C}}d\lambda(t)\,
e^{-\frac{1}{2}\left(\cc{t}^2+x^2\right)+x\cc{t}\sqrt{2}}F(t),
\ee
where the bar denotes complex conjugation. 

Specifically, the Bargmann transform of Hermite functions $\psi_n(x)$ is
\be
\mathcal{B}[\psi_n](t)=\frac{t^n}{\sqrt{n!}}.
\label{barg}
\ee
It has the algebraically important property that quantum numbers (state labels) $n$ add when single-particle wave functions are multiplied, $t^nt^m=t^{n+m}$. In this sense, replacing Hermite functions with Bargmann-space powers is analogous to replacing real standing waves $\cos kx$ with complex travelling waves $e^{ikx}$ (complexification), which have the same property with respect to their state labels, the wave numbers: $e^{ikx}e^{ik'x}=e^{i(k+k')x}$. Because $n!m!\neq (n+m)!$, we use unnormalized single-particle wave functions $t^n$, with scalar product
\be
(t^n,t^m)=n!\,\delta_{nm}.
\label{scalp}
\ee

In three dimensions, the Hermite functions in $x$, $y$, and $z$ are mapped onto Bargmann-space variables $t$, $u$, and $v$, respectively. The operators $x$ and $\partial_x$ in real space transform to
\be
x\bto \frac{1}{\sqrt{2}}(t+\partial_t),\quad
\partial_x\bto \frac{1}{\sqrt{2}}(-t+\partial_t)
\label{corr}
\ee
in Bargmann space, and similarly for the other two directions. Because the integral transforms factorize across the different space directions, the Bargmann transform of the angular momentum operator is obtained simply by inserting the transforms~\eqref{corr} into the usual expression:
\be
L_z=-i(x\partial_y-y\partial_x) \bto -i(t\partial_u-u\partial_t)
\equiv \mathfrak{L}_v,
\ee
and cyclically. Thus the form of the angular momentum operator is the same in real and Bargmann space. Therefore, solid harmonics in Bargmann space are the same polynomials as in real space, with $(x,y,z)$ simply replaced by $(t,u,v)$. Ladder operators are defined in the usual way. Because we work with unnormalized wave functions, the usual square root factor $\sqrt{(l+m)(l-m+1)}$  associated with lowering the projection is replaced by $(l+m)$.

\section{Results}

The general theory~\cite{Sunko16-1} gives the $2!^2=4$ shapes of two particles in three dimensions in the Cartesian basis:
\be
\Psi_1=t_1-t_2,\quad \Psi_2=u_1-u_2,\quad \Psi_3=v_1-v_2,\quad
\Psi_4=\Psi_1\Psi_2\Psi_3,
\label{basis}
\ee
where the indices in $t_1$ etc.\ refer to the two particles. The ground-state multiplet is evidently a vector $\vec{\Psi}=(\Psi_1,\Psi_2,\Psi_3)$, which may be written in spherical components:
\be
\Psi_{1,\pm 1}=\mp\Psi_1-i\Psi_2,\quad \Psi_{10}=\Psi_3.
\label{conshort}
\ee
The Euler boson basis~\cite{Sunko16-1} is given by the elementary symmetric functions~\cite{Stanley99}
\be
e_1(t)=t_1+t_2,\quad e_2(t)=t_1t_2,
\label{bos}
\ee
and similarly for $u$ and $v$. Because only $e_1$ carries one quantum of excitation, there are nine states in the first-excited shell, of the form $(a_1+a_2)(b_1-b_2)$ in the Cartesian basis, with $a,b=t,u,v$.

The Euler bosons $e_1$ may also be rewritten in the spherical basis, e.g.\ $e_{11}=-e_1(t)-ie_1(u)$. The only state with projection $m=2$ in the first-excited shell is evidently
\be
\Psi_{122}\sim e_{11}\Psi_{11},
\label{psi122}
\ee
where normalized wave functions are indexed with three quantum numbers, referring to the shell, $l$, and $m$, respectively. Because the lowering operator $\mathfrak{L}_-=\mathfrak{L}_t-i\mathfrak{L}_u$ (where $\vec{\mathfrak{L}} = \vec{\mathfrak{L}}_1 + \vec{\mathfrak{L}}_2$) is a linear combination of first derivatives, one sees immediately by the product rule that $\Psi_{121}\sim e_{10}\Psi_{11}+e_{11}\Psi_{10}$, therefore there exists an orthogonal state
\be
\Psi_{111}\sim e_{10}\Psi_{11}-e_{11}\Psi_{10},
\ee
which is, of course, part of a triplet, accounting for $8$ states so far. The ninth is the singlet
\be
\Psi_{100}\sim \vec{e}_1\cdot\vec{\Psi},
\ee
in obvious notation. In this way, all first-excited states are rewritten as bosonic excitations of the ground state, in a rotationally invariant manner. These states are called trivial because they are excitations of already known shapes, here the ground-state multiplet. The main interest is how the shape $\Psi_4$ appears in the second-excited shell in a rotationally invariant form, given that it is a pseudoscalar itself: it is the Bargmann-space image of the signed volume $(x_1-x_2)(y_1-y_2)(z_1-z_2)$.

The total number of states in the second shell is $28$. Two states with maximal projection $m=3$ may be constructed \emph{a priori}. One is chosen naturally according to Eq.~\eqref{psi122}:
\be
\Psi_{233}^{\RI}\sim e_{11}^2\Psi_{11},
\ee
while the other obvious choice,
\be
\Psi_{233}^{\RII}\sim \Psi_{11}^3,
\ee
turns out to be orthogonal to it. Because $\Psi_{233}^{\RI}$ is a product of different terms, while $\Psi_{233}^{\RII}$ is a pure cube, the product rule indicates that $\Psi_{232}^{\RI}$ is spanned by two vectors, $e_{11}e_{10}\Psi_{11}$ and $e_{11}^2\Psi_{10}$, while $\Psi_{232}^{\RII}$ consists of the one vector $\Psi_{11}^2\Psi_{10}$. Thus there is only one $l=2$ state, the one orthogonal to $\Psi_{232}^{\RI}$.

\begin{table*}
\begin{center}
\begin{math}
\begin{array}{ll}
\hline\\[-10pt]
\sqrt{128}\Psi_{233}^{\RI}=e_{11}^2\Psi_{11} &
\sqrt{192}\Psi_{232}^{\RI}=2e_{11}e_{10}\Psi_{11}+e_{11}^2\Psi_{10} \\
& \sqrt{1920}\Psi_{231}^{\RI}=2(2e_{10}^2 + e_{11}e_{1-1})\Psi_{11}
\\&\phantom{\sqrt{1920}\Psi_{231}^{\RI}=2}
 + 8e_{11}e_{10}\Psi_{10} + e_{11}^2\Psi_{1-1}
\\[2pt]\hline\\[-10pt]
\sqrt{384}\Psi_{233}^{\RII}=\Psi_{11}^3 & 
\sqrt{64}\Psi_{232}^{\RII}=\Psi_{11}^2\Psi_{10} \\ 
& \sqrt{640}\Psi_{231}^{\RII}= 4\Psi_{10}^2\Psi_{11} + \Psi_{11}^2\Psi_{1-1} \\[2pt]\hline\\[-10pt]
\sqrt{96}\Psi_{222}=e_{11}e_{10}\Psi_{11}-e_{11}^2\Psi_{10} & \sqrt{384}\Psi_{221}=(2e_{10}^2 + e_{11}e_{1-1})\Psi_{11}
\\&\phantom{\sqrt{384}\Psi_{221}=2}
- 2e_{11}e_{10}\Psi_{10} - e_{11}^2\Psi_{1-1}
\\[2pt]\hline\\[-10pt]
\multicolumn{2}{l}{
\sqrt{160}\Psi_{211}^{\RI}=\Psi_{10}^2\Psi_{11} - \Psi_{11}^2\Psi_{1-1}
}\\[2pt]\hline\\[-10pt]
\multicolumn{2}{l}{
\sqrt{384}\Psi_{211}^{\RII}=(-2e_{10}^2+e_{11}e_{1-1})\Psi_{11}+2e_{11}e_{10}\Psi_{10}-e_{11}^2\Psi_{1-1}
}\\[2pt]\hline\\[-10pt]
\multicolumn{2}{l}{
\sqrt{480}\Psi_{211}^{\RIII}=(e_{10}^2-2e_{11}e_{1-1})\Psi_{11}+2e_{11}e_{10}\Psi_{10}-e_{11}^2\Psi_{1-1}
}\\[2pt]\hline
\end{array}
\end{math}
\end{center}
\caption{Normalized wave functions $\Psi_{2lm}$ of the second-excited shell with angular momentum projection $m\ge 1$. Only the last two $l=1$ states are arbitrarily orthogonalized (to $\Psi_{231}^{\RI}$ and $\Psi_{221}$), because $\Psi_{211}^{\RI}$ is naturally orthogonal to $\Psi_{231}^{\RII}$ in the subspace of relative motion (see text).}
\label{wavefun}
\end{table*}

Further application of the product rule shows that the $m=1$ subspace is six-dimensional, spanned by the vectors $e_{10}^2\Psi_{11}$, $e_{11}e_{1-1}\Psi_{11}$, $e_{11}e_{10}\Psi_{10}$, and $e_{11}^2\Psi_{1-1}$, stemming from $\Psi_{232}^{\RI}$, and $\Psi_{10}^2\Psi_{11}$ and $\Psi_{11}^2\Psi_{1-1}$ from $\Psi_{232}^{\RII}$. Because three dimensions are already spoken for by two $l=3$ states and one $l=2$ state, there remain three $l=1$ states, for a total of $7+7+5+3+3+3=28$ states. There are no singlets: the product rule applied to the $m=1$ states gives only six $m=0$ states again (try it). All these states are given explicitly in Table~\ref{wavefun}.

Because of the scalar product~\eqref{scalp}, some care is required to obtain the orthonormalized wave functions in Table~\ref{wavefun}. We first calculate norms of all the monomials appearing in the table, such as $e_{11}e_{10}\Psi_{10}$, by expanding them in the underlying variables $t_1$ etc.\ and applying the formula \eqref{scalp}. (All these monomials are orthogonal to each other.) We then orthogonalize the vectors obtained with the lowering operator by applying Gaussian elimination to their scalar-product matrix. Finally the orthogonalized vectors are normalized using the previously calculated monomial norms.

It is easy to show that for \emph{normalized} states $\Psi_{23,\pm2}^{\RII}$ and $\widetilde{\Psi}_4=\Psi_4/\sqrt{8}$,
\be
\frac{1}{\sqrt{2}}\left(\Psi_{232}^{\RII}-\Psi_{23,-2}^{\RII}\right) = i\widetilde{\Psi}_4,
\ee
which means that the fourth shape is distributed equally among the $m=\pm 2$ states of $\Psi_{23m}^{\RII}$, and appears nowhere else. In other words, the signed volume $\Psi_4$ has good angular momentum $l=3$, albeit not good projection. Thus $\Psi_{23m}^{\RII}$ is the only non-trivial multiplet in the second shell, and all states in the spectrum can be obtained in the form of a superposition of bosonic excitations of the ground state and of this state. Hence the spectrum of two identical fermions in an oscillator well is naturally split into two bands, with $\Psi_{01m}$ and $\Psi_{23m}^{\RII}$ as their respective band-heads. This exact kinematic result is paradigmatic for all finite systems of fermions.

\section{States of relative motion}

Removing the center-of-mass (CM) motion from a two-body problem reduces it to a one-body problem. It is important to establish that the above classification in two bands is relevant for states of relative motion (RM) by themselves. The Bargmann transform of $x_1\pm x_2$ is $t_1\pm t_2$. Writing $t=t_1-t_2$ etc., RM can be expressed by a triplet of single-particle variables $(t,u,v)$ in Bargmann space.

The Bargmann transform~\eqref{barg} of an odd function in real space is an odd function in Bargmann space. The CM wave function is even, so the RM part must change sign when $(t,u,v)\rightarrow (-t,-u,-v)$. A general RM wave function in Bargmann space then reads
\be
Pt+Qu+Rv+Stuv,
\ee
where $P$, $Q$, $R$, $S$ are functions of squares $(t^2,u^2,v^2)$ \emph{only} (noting that $P=tu$ is the same as $Q=t^2$). One recognizes the constraint~\eqref{scheme} in the basis~\eqref{basis}, obtained by inspection here.

Therefore, excitations of RM must be expressed in discriminants,
\be
\Delta_2(t)\equiv (t_1-t_2)^2=e_1(t)^2-4e_2(t),\quad \Delta_2(u),\quad 
\Delta_2(v),
\ee
so all excitations which are not expressed in this way are CM motions. These are, in particular, all powers of the Euler bosons $e_1$ appearing without $e_2$ in the above combination. Evidently, $e_2=(e_1^2-\Delta_2)/4$ is redundant, so a more physical basis for the bosonic excitations than the textbook symmetric functions~\eqref{bos} is
\be
e_1(t)=t_1+t_2,\quad \Delta_2(t)=(t_1-t_2)^2,
\label{disc}
\ee
the first boson being responsible for CM excitations, and the second for RM ones.

Hence most of the states in Table~\ref{wavefun} contain CM excitations. The only pure-RM states are the septiplet $\Psi_{23m}^{\RII}$ and triplet $\Psi_{21m}^{\RI}$. (They correspond to single-particle 3D-oscillator functions with three quanta, of which there are ten, as easily checked.) Now observe that  $\Psi_{23m}^{\RII}$ contains the pseudoscalar $\Psi_4$, while $\Psi_{21m}^{\RI}$ does not. Thus there are two distinct bands by symmetry in RM alone, QED. We are in a position to characterize them physically now.

$\Psi_{23m}^{\RII}$ is the lowest state with $l=3$, i.e.\ just the Bargmann transform of the solid harmonic $r^3Y_3^m$, while $\Psi_{21m}^{\RI}$ is the first excited state with $l=1$, so it contains $n_r=1$ quantum of radial excitation ($2n_r+l=3$). The orthogonalization which gave $\Psi_{21m}^{\RI}$ thus corresponds to replacing angular with radial quanta. Because even powers of $\Psi_4$ are discriminants, odd powers contain $\Psi_4$ linearly. There is an infinite number of excited RM states both with and without $\Psi_4$: neither band is terminated kinematically. The qualitative conclusion, in time-honored language, is that the band containing $\Psi_4$ is rotational, while the other one is vibrational.

Notably, the factor $2$ in $2n_r+l$ is due to the discriminant in Eq.~\eqref{disc} being a square. Thus the well-known result that each radial quantum carries two oscillator quanta can be understood without reference to an equation of motion. Such reasoning reflects the kinematic (``off-shell'') nature of the constraint~\eqref{scheme}.

\section{Discussion}

Because of particle indistinguishability, the phase space available for interactions is highly restricted in the vicinity of the ground state. The Fermi-liquid idea is that when interactions are turned on adiabatically, the ground-state wave function evolves smoothly, keeping the form of a ground-state Slater determinant~\cite{Mahan90}. The interactions only affect the composition of the quasiparticles which make up this determinant: under adiabatic conditions, there is not enough phase space for anything else. New ground states and collective excitations appear by coherent accumulation of small perturbations of the Slater-determinant ground state, typically particle-hole excitations, following Bloch's idea.

The Fermi-liquid paradigm corresponds to ``linear combination of atomic orbitals'' (LCAO) in quantum chemistry. An early alternative to it for the description of strongly correlated states in finite systems was the Heitler-London wave function~\cite{Mattis65}. The more recent configuration-interaction approach~\cite{Bressanini12} explicitly begins with a superposition of Slater determinants with roughly equal amplitude, in an attempt to control some chosen correlation from the outset. Such wave functions are non-perturbative with respect to the Fermi-liquid ground state. They are naturally competitive at least in finite systems, where the quasiparticle (LCAO) construction is itself subject to gapping because of their discrete spectrum. The same restriction of phase space at low energy applies to them as well, however, opening the question how to choose this superposition judiciously. One does not want to introduce ``trivial'' components in a ground-state \emph{ansatz}, which cost energy without contributing correlations.

The present approach gives a precise kinematic prescription, which states are trivial: those which contain Euler bosons. Explicitly, non-trivial states are those which can be expressed with $\Phi_i\in\mathbb{C}$ in Eq.~\eqref{scheme}. The prescription depends on a special class of antisymmetric wave functions, the shapes $\Psi_i$, which reconstruct the whole Hilbert space as a free module, with symmetric-function coefficients. An algorithm to generate the shapes has been published~\cite{Sunko16-1}.

The free-module structure of Eq.~\eqref{scheme} is a direct connection between fundamental quantum mechanics and algebraic geometry. Algebraic geometry relates nodal (zero) surfaces of many-body wave functions to polynomial ideals. Shapes are geometric objects in wave-function space, which constrain fermionic motion kinematically according to Eq.~\eqref{scheme}. These geometric constraints are the source of strong correlations. The connection with algebraic geometry provides the particle side of the particle/field dichotomy with a mathematical framework as rich as differential geometry provides for field theory.

This algebraic structure qualitatively resolves a foundational problem in the physics of finite systems, most notably discussed in nuclear physics~\cite{Bohr53,Bohr76}. The problem is why spectra are organized into bands: classes of excited states with similar properties, and dissimilar between the classes. The reason is kinematic: the band-head of each band is  qualitatively different from the other band-heads, to which it cannot be connected by a bosonic excitation. As shown in the elementary example here, even at two-particle level there appears one additional vacuum shape $\Psi_4$, which is not a single Slater determinant like the ground state $\vec{\Psi}$. There are two bands because there are two shapes constraining the motion.

One may well ask how such a fundamental structure of Hilbert space could escape the attention of physicists, to the extent that the underlying theory of algebraic invariants~\cite{Derksen02} is not even part of the standard physics curriculum, although it was completed by Hilbert already in the 1890's.

To be fair, a particular example of it has been observed in the context of the $d=2$ fractional quantum Hall effect (FQHE), albeit without noticing its generality. Namely, Laughlin's $N=3$ wave function for the FQHE, Eq.~(7.2.12) of Ref.~\cite{Laughlin90}, can be factored as $\mathcal{N}\Phi_m\Phi_n\Psi_0\mathcal{E}$, where $\mathcal{N}$ is a norm, $\mathcal{E}$ an exponential localization term, and, in the notation of Ref.~\cite{Laughlin90},
\be
\Psi_0=(z_1-z_2)(z_1-z_3)(z_2-z_3)=
\left|
\begin{matrix}
z_1^2 & z_2^2 & z_3^2\\
z_1 & z_2 & z_3\\
1 & 1 & 1
\end{matrix}
\right|
\ee
is the only linear combination of shapes with three nodes, among the four found in Ref.~\cite{Sunko16-1} for the case $N=3$, $d=2$, which can be written in terms of the $z_i$ alone. The other three involve $\bar{z}_i$ on one or both rows of the determinant, while two other shapes, with two and four nodes~\cite{Sunko16-1}, similarly involve both $z_i$ and $\bar{z}_i$, for a total of $N!^{d-1}=6$ shapes. (The $\Phi_{m,n}$ may be rewritten in elementary symmetric functions of the $z_i$, called Euler bosons here.)

The above argument is a formal proof (by enumeration!) of Laughlin's conjecture that there are no other vacua with $N=3$ and $d=2$ which satisfy the analyticity constraint. Given the large body of research devoted to this special case, it is even more curious that its general aspects remained unnoticed. There are two disadvantages of wave functions in real space which may be relevant to this issue, although it should also be noted that mathematicians have begun to study the specifics of Hilbert's invariant theory in three dimensions only recently, apparently without being aware of the physics connection~\cite{Bergeron12}.

First, in real space, quantum numbers are indices of special functions, which have opaque multiplication properties. To see the real-space counterparts of perfectly transparent expressions such as $e_{11}^2\Psi_{10}$, one has to multiply them out, ``flattening'' the algebraic structure, and then replace each power of $t$, $u$, and $v$ with the corresponding (unnormalized) Hermite function. Such a procedure typically gives a superposition of Slater determinants which appears quite unmotivated.

Second, real one-body wave functions oscillate between first-order zeros merely to be orthogonal, while hiding the non-trivial Pauli zeros of the many-body function in interference effects~\cite{Ceperley91}. In Bargmann space, trivial one-body zeros collapse to the $n$-fold zeros of monomials $t^n$~\eqref{barg} at the origin, while the physical many-body nodal surfaces are zeros of the corresponding polynomials~\eqref{scheme}. Thus complexification avoids both disadvantages of real wave functions, providing the natural setting for the multiplicative free-module structure to be observed.

To the above structural problems with real-space wave functions, one may add the problem with the norm, mentioned previously. Because $n!m!\neq (m+n)!$, using normalized one-particle wave functions hides the algebraic structure behind a trivial technicality.

Importantly, shapes cannot contain any Euler-boson zero-point energy, because they are \emph{defined} as the orthogonal complement to states which contain Euler bosons~\cite{Sunko16-1}. The excitations of the shapes are always decomposable in elementary symmetric functions, i.e.\ may be considered bosonic, whether they are collective from the physical point of view or not. The fermion content of these bosons is manifest: they are symmetric polynomials in the same one-body wave functions as the vacua, whose wave functions (shapes) are antisymmetric in them. The decomposition~\eqref{scheme} avoids double counting zero-point energy by construction.

These general considerations should not obscure the simplicity with which the product rule was harnessed on the multiplicative structure to reduce the 28-dimensional space of the second-excited shell into angular-momentum multiplets. There is hardly a simpler way to discover that $28=7+7+5+3+3+3$, and find the spanning vectors. Because Euler bosons have physical meaning, distinct strings of quantum numbers in different monomials such as $e_{11}e_{10}\Psi_{10}$ and $e_{11}^2\Psi_{1-1}$ guarantee them to be distinct vectors in Hilbert space, indeed orthogonal in this case. For this reason, it is easy to find the dimensions of the invariant subspaces by ``counting on fingers'' with the help of the product rule.

The present example shows both possible ways in which shapes can appear within a rotationally invariant scheme. One is that some set of pure shapes is an angular-momentum multiplet by itself, such as the ground-state vector $(\Psi_1,\Psi_2,\Psi_3)$ here. The other is that a shape, like $\Psi_4$ here, appears in linear superposition with trivial states --- bosonic excitations of lower shapes --- to form a state with good $l$ and $m$. Superposition with trivial states must always be possible to fix both $l$ and $m$, because each shell by itself can be decomposed into states of good angular momentum. Because shapes are orthogonal to the trivial states, it follows that the shape subspace itself can always be resolved into shapes of good angular momentum, although not necessarily good $m$-projection. A resolution of the shape subspace into states of good angular momentum is an orthogonal projection of some such resolution of the complete shell.

A special feature of this small example is that the components of $\vec{\Psi}=(\Psi_1,\Psi_2,\Psi_3)$ divide $\Psi_4=\Psi_1\Psi_2\Psi_3$ as polynomials, which happens sporadically in general. Nothing in the above reasoning depended on it. The pure cube $\Psi_{233}^{\RII}$ can be resolved in Euler bosons ultimately because squares of the $\Psi_i$ are discriminants, so that it is an Euler boson excitation of $\vec{\Psi}$, despite appearances. On the other hand, the symmetric function $\Psi_1\Psi_2$ cannot be resolved in Euler bosons, so $\Psi_4$ is a shape, not a bosonic excitation of $\Psi_3$ in this precise sense.\footnote{The Euler bosons are a complete basis of symmetric functions in each direction separately, while $\Psi_1\Psi_2$ factorizes into \emph{antisymmetric} functions across the two directions.} Therefore $\Psi_{232}^{\RII}$ is an example of the general scheme~\eqref{scheme} with $D=4$: a superposition of \emph{different} vacua, $\vec{\Psi}$ and $\Psi_4$, adjusted to have good angular momentum and projection.

The natural initial choices $e_{11}^2\Psi_{11}$ and $\Psi_{11}^3$ for $m=3$ states need not be as obvious in a problem with more particles. The initial choice of maximum-projection states does not affect the shapes themselves, which form a closed subspace in each shell~\cite{Sunko16-1}. However, it does affect the optimal choice of basis in the shape subspace, and the precise way in which shapes are embedded in the multiplets containing them.

To conclude, we have given a detailed textbook exposition of the simplest possible case of exact algebraic bosonisation for $N=2$ particles in $d=3$ dimensions. The spectrum of two particles in three dimensions is kinematically classified in two distinct bands, one the ground-state band, and the other with band-head in the second oscillator shell. The classification is paradigmatic for all rotationally invariant finite systems.

\section{Acknowledgements}

This work was supported by the Croatian Science Foundation under Project No.~IP-2018-01-7828 and University of Zagreb Support Grant 20283207. KR thanks the Croatian Physical Society for a travel grant.

\end{document}